\documentclass[pra,superscriptaddress,longbibliography, twocolumn,amsmath,amssymb]{revtex4-1}	
\usepackage{graphicx}
\usepackage{dcolumn} 
\usepackage{hyperref}
\hypersetup{
    colorlinks=true,        
    linkcolor=blue,          
    citecolor=blue,        
    filecolor=magenta,      
    urlcolor=blue           
}
\newcommand{\Figref}[1]{Fig.\ref{#1}}
\newcommand{\bR}{\mathbf{R}}
\newcommand{\citenamefont}{}
\newcommand{\bibnamefont}{}

\usepackage{color} 				
\usepackage[normalem]{ulem} 			

\begin{document}
\title{The Dielectric Wakefield Resonator Accelerator}

\author{G.V. Sotnikov}\email{sotnikov@kipt.kharkov.ua; Gennadiy.Sotnikov@yandex.ua}
\affiliation{NSC Kharkov Institute of Physics and Technology, 61108 Kharkov, Ukraine}
\author{T.C. Marshall}
\affiliation{Columbia University, 10027 New York city, USA}
\author{S.V. Shchelkunov}
\affiliation{Yale University, 06520 New Haven, CT, USA}
\author{J.L. Hirshfield}
\affiliation{Yale University, 06520 New Haven, CT, USA}
\affiliation{Omega-P, Inc., 06511 New Haven CT, USA}

\date{\today}

\begin{abstract}
  We report preliminary studies of a three-channel, rectangular, high gradient dielectric wakefield accelerator element, which, unlike the collinear cylindrical dielectric wakefield device, does not suffer from low transformer ratio and may offer relief from the beam breakup instability.  When configured as a resonator (DWR), it can be driven by a series of modest-charge drive bunches. This rectangular ‘mode-locked’ resonator consists of three channels lined with low-loss dielectric slabs: the central wider channel is the drive bunch channel, whereas two adjacent narrow channels are used to accelerate electron and/or positron bunches; this provides a favorable transformer ratio. At the moment when, after reflecting from the resonator exit, the wakefield returns to the resonator entrance, the next bunch is injected into the resonator entrance. The length of the resonator is also chosen to be a multiple of half the desired wakefield wavelength. The rectangular geometry permits superposition of harmonics of wakefields in the resonator. The short length of the device and its planar configuration should allow management of beam breakup, and a lengthy drive bunch train can be dynamically stabilized by using a series of DWR units rotated about the $z$-axis in 90° increments.
\end{abstract}
\maketitle

\section{Introduction}
Dielectric-lined structures show promise for generating strong accelerating fields arising from relativistic electron bunches (DWA) \cite{Gai1988PRL,Ros1990PRD,*Ng990PRD,Park2000PRE,Power2000PRSTAB}.  Most of the studies on the use of dielectric structures have been carried out for cylindrical collinear configurations, but achieving both high accelerating gradient and elevated transformer ratio in this configuration appears to be impossible~\cite{Baturin2017PRSTAB}. Attention also has been directed to dielectric-lined waveguides having rectangular configuration \cite{Zha1997PRE,Xiao2001PRE,Mar2001PRSTAB,Mar2002AAC-10,Wang2004PRSTAB,Yoder2005PRSTAB,Shchelkunov2012PRSTAB,Wang2006PRSTAB}. Simplicity of manufacturing, possibility of realizing a multimode regime of excitation with equally-spaced frequencies \cite{Zha1997PRE} resulting in a significant increase of accelerating field amplitude, easy fine tuning of working frequency, additional intrinsic focusing in an accelerating field~\cite{Xiao2001PRE}, availability of low-loss, dispersion-free dielectric materials, make dielectric-lined structures in a rectangular configuration attractive for excitation of accelerating fields by a laser pulse \cite{Yoder2005PRSTAB} or electron bunches \cite{Zha1997PRE,Xiao2001PRE,Mar2001PRSTAB,Mar2002AAC-10,Wang2004PRSTAB, Shchelkunov2012PRSTAB,Wang2006PRSTAB}.

The excitation of a single-channel rectangular waveguide resonator lined with two symmetric dielectric slabs by a uniformly spaced train of electron drive bunches has been studied~\cite{Xiao2001PRE,Mar2002AAC-10,Wang2006PRSTAB}.  The separation of waves into LSM and LSE modes is very effective for such problems~\cite{Pincherle1944PR}. Each of the family of LSM and LSE modes, in its turn, contains odd and even symmetric modes. On the basis of analytical expressions, a numerical analysis of wakefield excitation in the resonator for the particular case when the frequency of bunch repetition coincides with the frequency of the fundamental odd LSM mode or the fundamental odd LSE mode has been carried out~\cite{Sot2006AAC}.

We now study a symmetric, rectangular dielectric loaded unit consisting of a central channel bordered by two narrow side channels; four slabs of dielectric are used, as in Fig. 1.  One or more drive bunches can traverse the central channel and thereby set up intense wakefields that will accelerate witness bunches located in the narrow side channels.  This device is essentially a symmetrized version of a two-beam rectangular DWA~\cite{Shchelkunov2012PRSTAB} that has been operated successfully.

\begin{figure}
  \centering
  \includegraphics[width=0.5\textwidth]{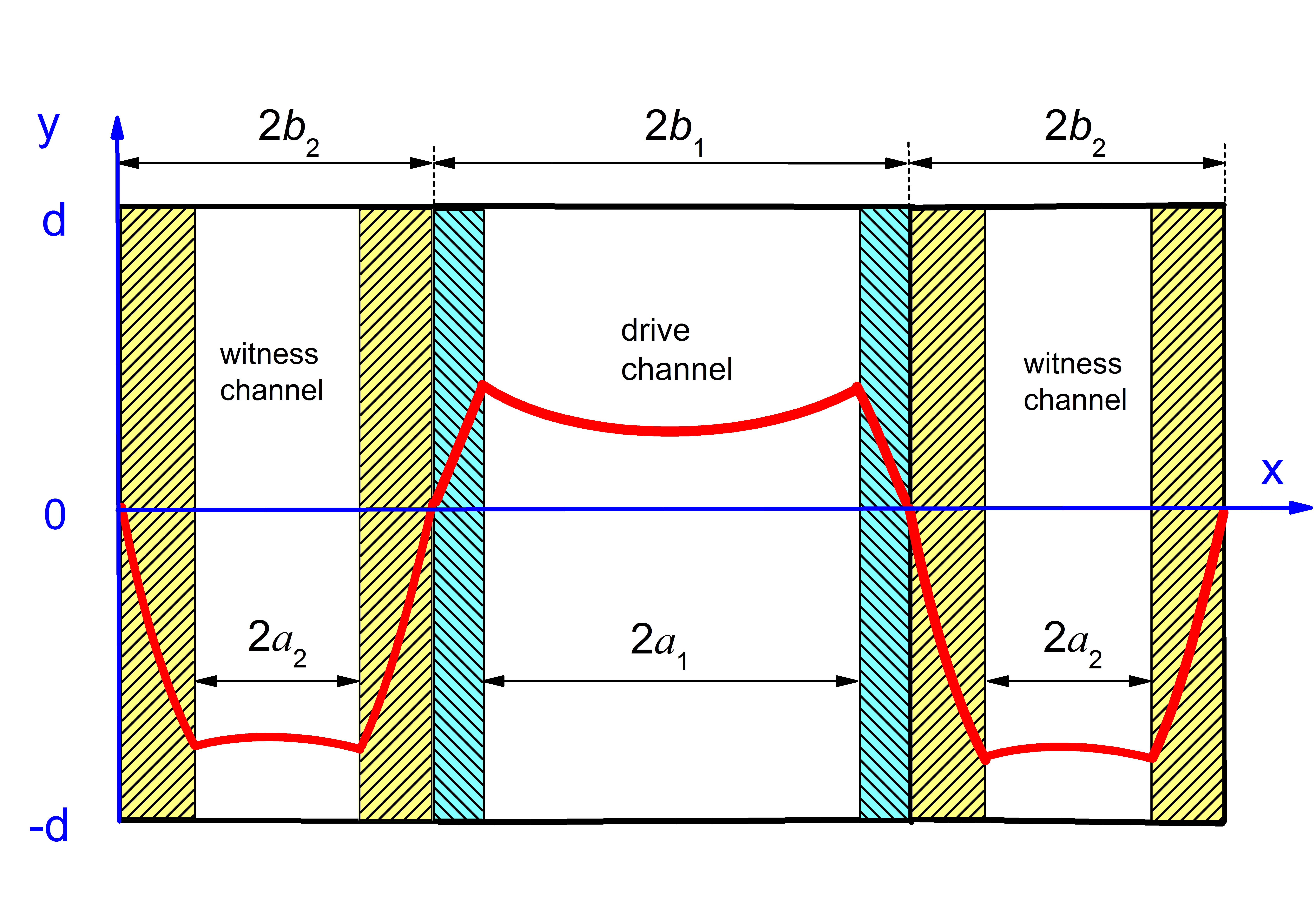}
  \caption{Schematic section of a 3-channel DWA, showing arrangement of drive and witness bunch channels, dielectric slabs (cross-hatched), and the wakefield $E_z$ profile (red) of the $LSM_{41}$  mode set up by the drive bunch.  Dielectrics may differ in the drive and witness channels.}\label{Fig:01}
\end{figure}

This accelerating module will have an elevated transformer ratio $\bR$ if the side channels are narrower than the drive channel (\Figref{Fig:01}) and the $LSM_{41}$ mode, as shown, is excited.  Separation of drive and witness channels allows achieving both high accelerating gradient and elevated transformer ratio $\bR$.  The requirement $\bR \gg 2$ is needed for efficient operation of a dielectric wakefield high energy accelerator system, to minimize the number of drive beam segments needed to achieve a given final test beam energy; the cost of such an accelerator tends to scale as $(\bR)^{-1/2}$ \cite{Shiltsev2015DPF,Shchelkunov2015EAAC}.  The geometry we have chosen to study also offers an opportunity to deal with the problem of the beam breakup (BBU) instability \cite{Li2014PRSTAB} which will limit the gradient that can be obtained in a collinear DWA.   If it is “tall”, a rectangular DWA structure excited by a “sheet” bunch has weaker transverse wakefields (none in the 2D limit) than the cylindrical DWA \cite{Tremaine1997PRE}, thereby providing a certain remediation of the BBU.  A sheet-type witness bunch configuration (also referred to as “flat” or “pancake”) is also potentially useful in a Collider.

\begin{figure*}
  \centering
  \includegraphics[width=0.75\textwidth]{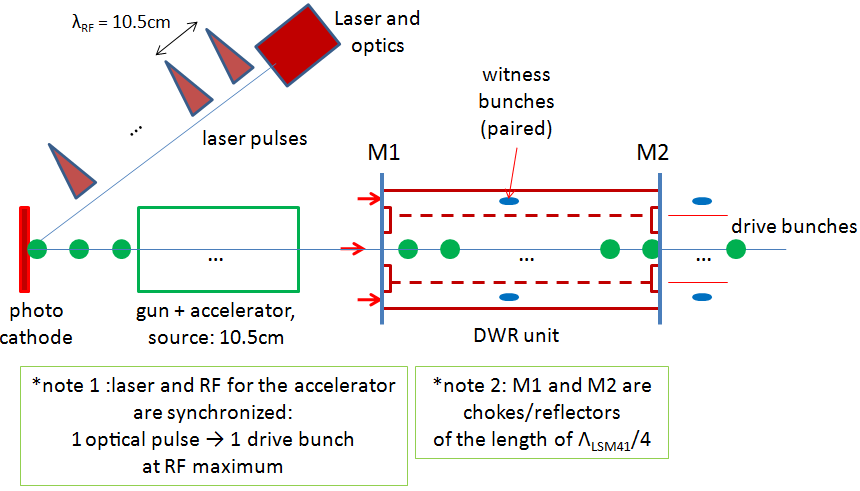}
  \caption{DWR excited by a train of drive bunches (green) caused by a series of optical pulses from a laser (red) that are synchronized to the RF field in a linac; a pair of witness bunches (blue) trails behind a drive bunch that moves through the resonator when the wakefield there has maximized.}\label{Fig:02}
\end{figure*}

We begin by describing a rectangular three-channel DWA unit that can accommodate sheet-type bunches.  It can be excited by a single high-charge drive bunch.  Following that is a discussion of how the performance of this unit might be improved if a section of it is enclosed by reflectors, which we then refer to as a dielectric wakefield resonator (DWR) \cite{Mar2001AAC-9,Sot2006AAC,Kis2006AAC}.  This DWR unit can be excited by a train of low-charge drive bunches that will cause the accumulation of an intense wakefield inside the resonator structure.  At the moment when, after reflecting from the resonator exit, the wakefield returns to the resonator entrance, the next bunch is injected into the resonator entrance. The length of the resonator is also chosen to be a multiple of half the desired wakefield wavelength by design of the dielectric liners.  \Figref{Fig:02} is a schematic showing how the DWR might be used as a component of a high energy wakefield linear accelerator.  The train of drive bunches is made by a macropulse of laser pulses, incident upon a photocathode, that occur near the peaks of the RF field in the gun and the drive bunch accelerator system. The advantage of the DWR is that a train of drive bunches spaced so as to enhance the $LSM_{41}$ wakefield by coherent superposition will suppress unwanted wakefield modes which can decrease R; using multiple bunches to build up the wakefield also may present less demand on the injection RF gun and accelerator.  Examples follow in the next two sections.

\section{The Three-Channel Rectangular DWA}
Small dimensions of the unit will favor high gradients, so as an example, we consider a unit with the following specifications (Table~\ref{tab:01}):
\begin{table}
\centering
\caption{Parameters for a three-beam channel structure (Cordierite)}\label{tab:01}
\begin{tabular}{|l|r|}
\hline
frequency of $LSM_{41}$ design mode          & $285.2\,GHz$         \\
accl. channel dimensions $2a_2$              & $0.2\,mm$            \\
drive channel dimensions $2a_1$              & $2.4\,mm$            \\
structure height $2d$                        & $6\,mm$              \\
slab-1 thickness                             & $0.115\,mm$          \\
slab-2 thickness                             & $0.137\,mm$          \\
slab-3 thickness                             & $0.137\,mm$          \\
slab-4 thickness                             & $0.115\,mm$          \\
slab relative dielectric constant            & $4.76$               \\
bunch dimensions $x_b\times y_b \times z_b$  &                      \\
(box distribution)                           & $0.1\times 0.1 \times 0.1\,mm^3$ \\
drive bunch energy                           & $200\,MeV$             \\
drive bunch charge                           & $2.5\,nC$            \\
drive bunch number                           & $1$                  \\
drive bunch center location, $x_{dr}$        & $1.622\,mm$          \\
\hline
\end{tabular}
\end{table}

\begin{figure}
  \centering
  \includegraphics[width=0.5\textwidth]{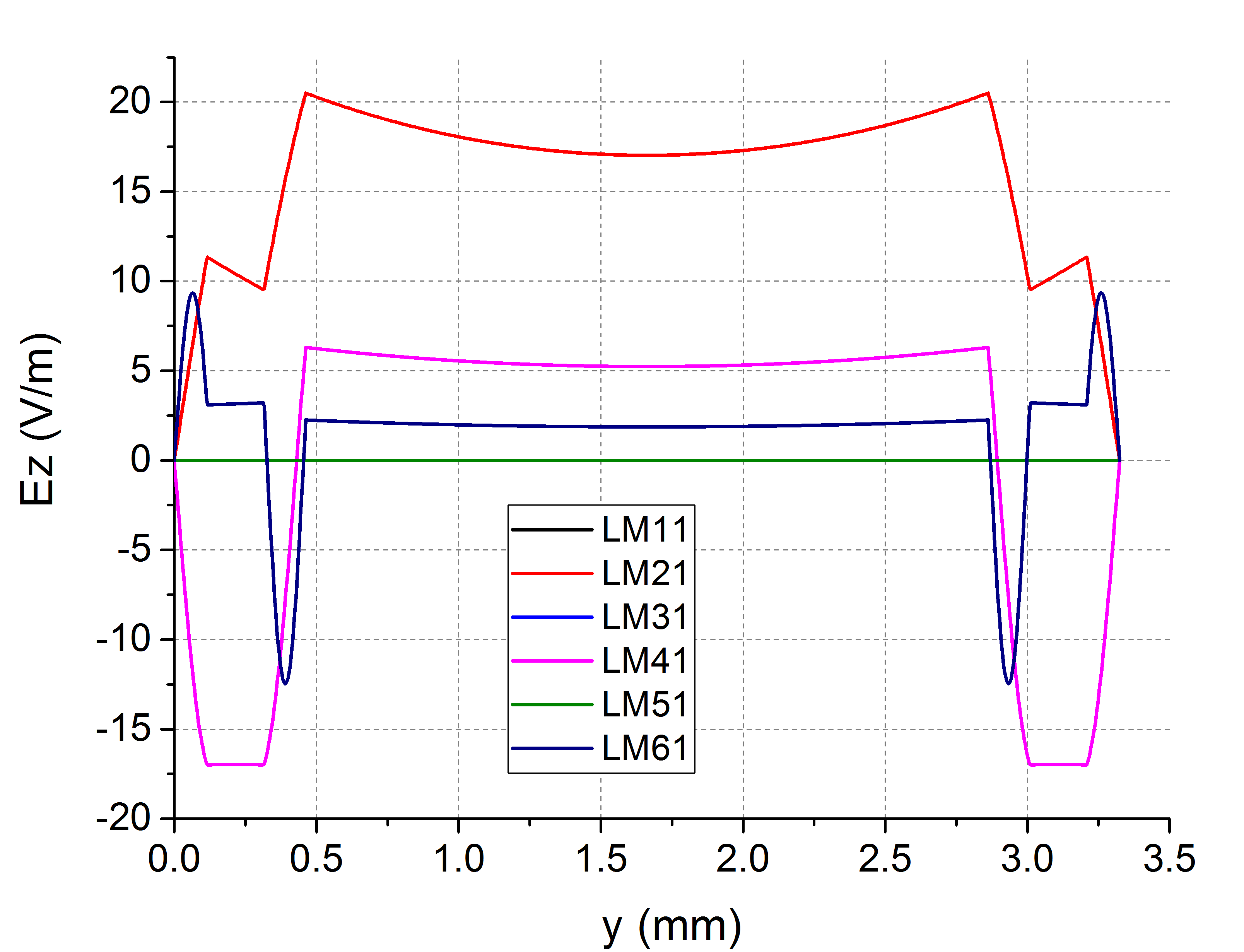}
  \caption{Axial wakefield $E_z$ vs $x$ for the acceleration peak of each mode.  Number of drive bunches is 1}\label{Fig:03}
\end{figure}

The wavelengths of the modes $LSM_{mn}$ ($m$ is the $x$-subscript, $n$ is the $y$-subscript), $mn=11; 21; 31; 41; 51; 61$ are as follows: $\lambda_{mn}= 0.4986; 0.3371; 0.1127; 0.1051; 0.0512; 0.05\,cm$.  Profiles of some of these modes are shown in \Figref{Fig:03}.  The $LSM_{41}$ has a desirable $\bR$ whereas the other modes do not.

\begin{figure}
  \centering
  \includegraphics[width=0.5\textwidth]{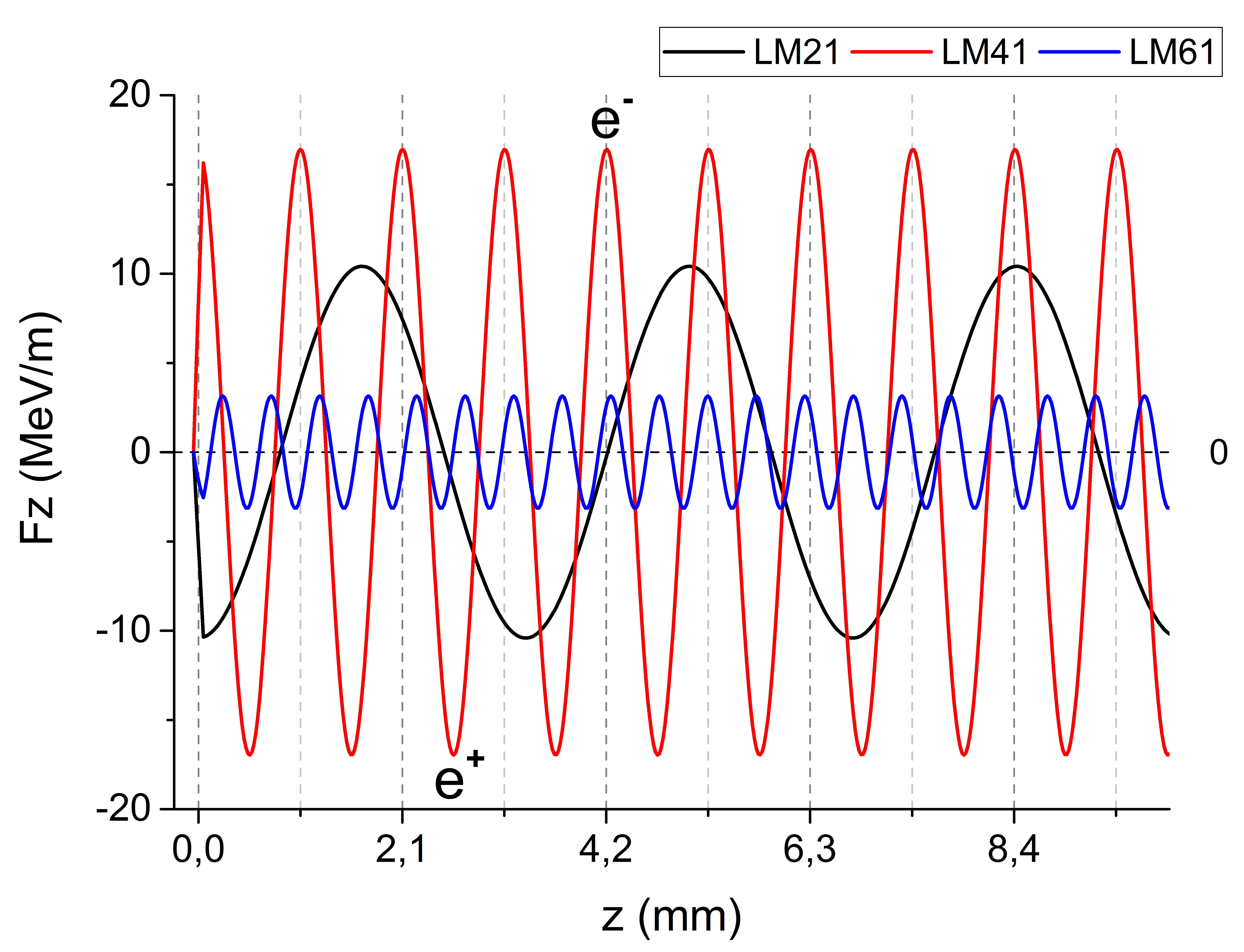}
  \caption{Axial wakefield force at the center of a witness bunch channel from a single $2.5\,nC$ drive bunch moving in the central channel.  Black line is the $LSM_{21}$ component; red line is $LSM_{41}$; blue line is $LSM_{61}$.  Very little energy goes into $LSE$ modes.  Position $e^-$ or $e^+$ indicates where an electron or positron witness bunch should be placed to interact primarily with the $LSM_{41}$ mode.  }\label{Fig:04}
\end{figure}

To determine the behavior of this structure, we, using the analytical theory for excitation of multi-zone dielectric structures by electron bunches~\cite{Wang2006PRSTAB,Sot2008PAST-NPI}, have made analytical computations of the wakefield excitation in a waveguide with the cross-section given by Table~\ref{tab:01}, set up by the passage of a single “point-like” drive bunch as specified therein.  The result is shown in \Figref{Fig:04}.  This electron bunch excites only modes with symmetric transverse profile of the $E_z$ in the drive channel. The desired operating mode with symmetric $E_z$ profile in all three channels is $LSM_{41}$ which has a frequency of $285.2\,GHz$.  For this single $2.5\,nC$ bunch, analytic computation finds that an axial wakefield force on a test electron of $17\,MeV/m$ appears in the witness channel, achieving $\bR = 6.6$ computed for the average decelerating force acting upon the drive bunch.  \Figref{Fig:04} shows that for this design there is competition between the desired $LSM_{41}$ mode and the unwanted $LSM_{21}$ mode.  However, a witness bunch can be located at a position where the $LSM_{41}$ mode amplitude dominates the others, as shown in the figure at axial positions $e^{+}/e^{-}$. Although modes competing with $LSM_{41}$ can be reduced in importance by using a long train of drive bunches, the trailing bunches can be destabilized by head-to-tail deflections after moving a a certain distance: this problem motivates the use of the DWR.

\section{The Three-Channel Rectangular Dielectric Wakefield Resonator (DWR)}

The use of cavity resonators to accelerate particles is well-established physics.  By injecting several drive bunches spaced by a multiple of the desired $LSM_{41}$ mode and reflecting the wakefield radiation from tuned mirrors, we can suppress the buildup of unwanted non-resonant radiation in a DWR.  Given favorable resonator $Q$, fields can accumulate in the compact resonator section to high amplitude.  The theory and experimental operation of a single-channel DWR was reported several years ago \cite{Sot2006AAC,Kis2006AAC,Oni2006AAC-12}.  If the accelerator providing the drive bunch train were to be powered by a $10.5\,cm$ RF source, the length of a resonator that holds one bunch would be $10.5\,cm$ and include $100$ resonant periods of the $LSM_{41}$ mode.

\begin{figure}[!tbh]
  \centering
  \includegraphics[width=0.48\textwidth]{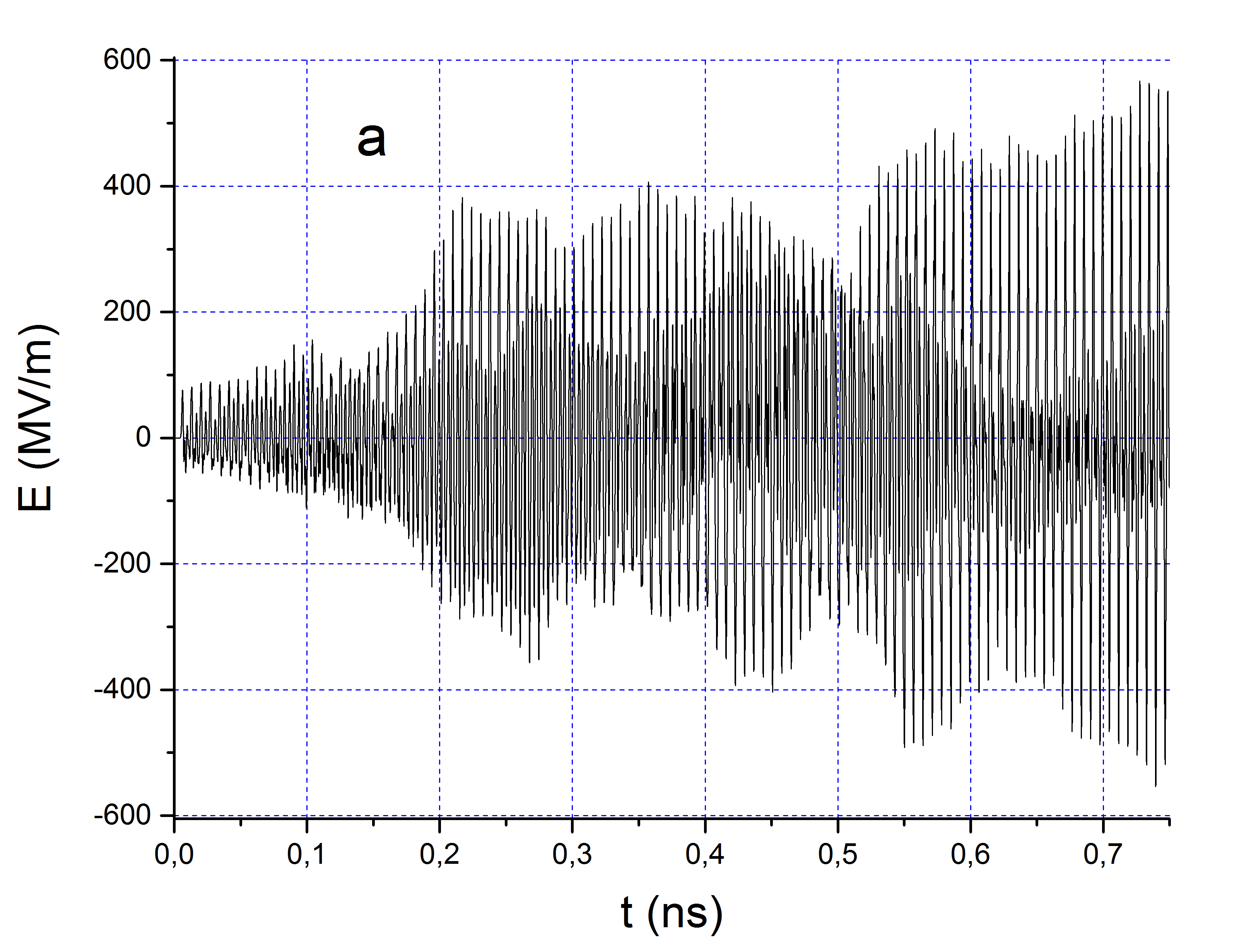}\hfill
   \includegraphics[width=0.48\textwidth]{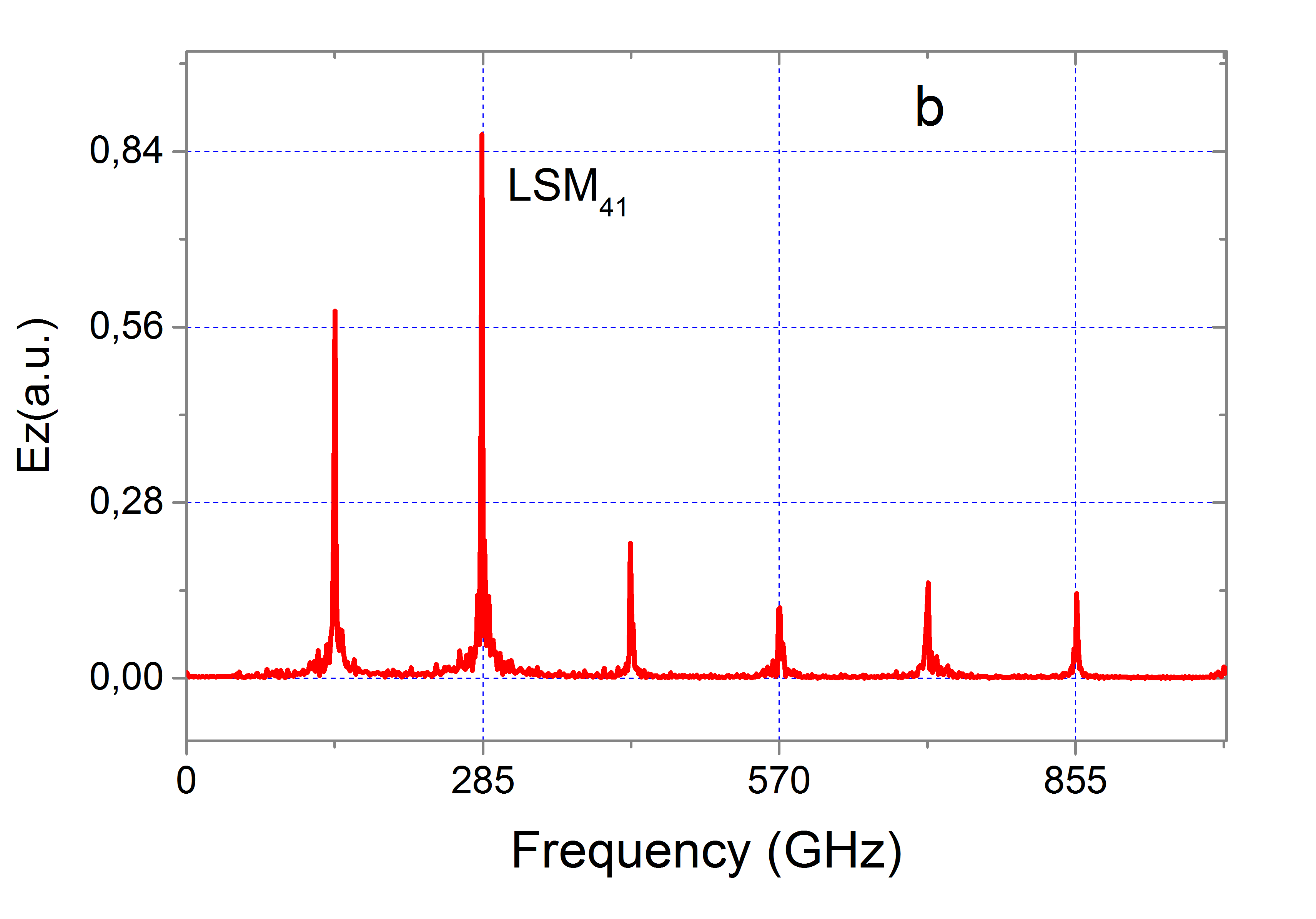}
  \caption{PIC simulations obtained using CST Studio: axial wakefield amplitude (above, \Figref{Fig:05}a) and its spectrum (below, \Figref{Fig:05}b) at the input of the witness channel. The model three channel resonator is excited by a train of $40$ $2.5\,nC$ bunches. The resonator length is $10.5\,mm$ ($10$ wavelengths of the $LSM_{41}$  mode), and the bunch spacing is $2.1\,mm$ ($2$ wavelengths of the $LSM_{41}$  mode).  The darker shaded regions in \Figref{Fig:05}a are caused by modes competing with the $LSM_{21}$ , as shown in \Figref{Fig:05}b.}\label{Fig:05}
\end{figure}

To study the performance of a three channel device configured as a resonator we use the CST Studio code, but because the time required for computation of a unit that would be suitable in practice is too large, a model structure was tested using a tight bunch spacing in a short resonator. A PIC simulation taking the resonator length to be $L = 10.5\,mm$ (to hold $5$ bunches at a time) with bunch spacing of $2.1\,mm$ (two $LSM_{41}$ wakefield periods) is shown in \Figref{Fig:05}.  The resonator is therefore $20$ half-wavelengths of the desired mode in length. A peak accelerating field in the resonator $≥500\,MeV/m$ is obtained at its input (\Figref{Fig:05}a) after injecting $40$ $2.5\,nC$ bunches.  The wakefield spectrum (\Figref{Fig:05}b) shows the $LSM_{41}$ mode  prevails for this shortened device, so we can expect  to obtain high transformer ratio here, as was the case for the simple waveguide computation (competing modes with appreciable power would lower the $\bR$ because of unfavorable profiles). The $LSM_{41}$  mode is preferentially excited because of the destructive interference of the non-resonant modes in the designed resonator.  The spectrum peak to the left of the $LSM_{41}$  mode is an artifact that is caused by the choice of drive bunch spacing of twice the $LSM_{41}$  period.  The field within the resonator can be represented as two oppositely-directed interfering waves, so the gradient available for accelerating a given bunch is one-half the gradient of the standing wave.  The witness bunch is to be accelerated by the co-moving component of the standing wave wakefield.

\begin{figure}
  \centering
  \includegraphics[width=0.5\textwidth]{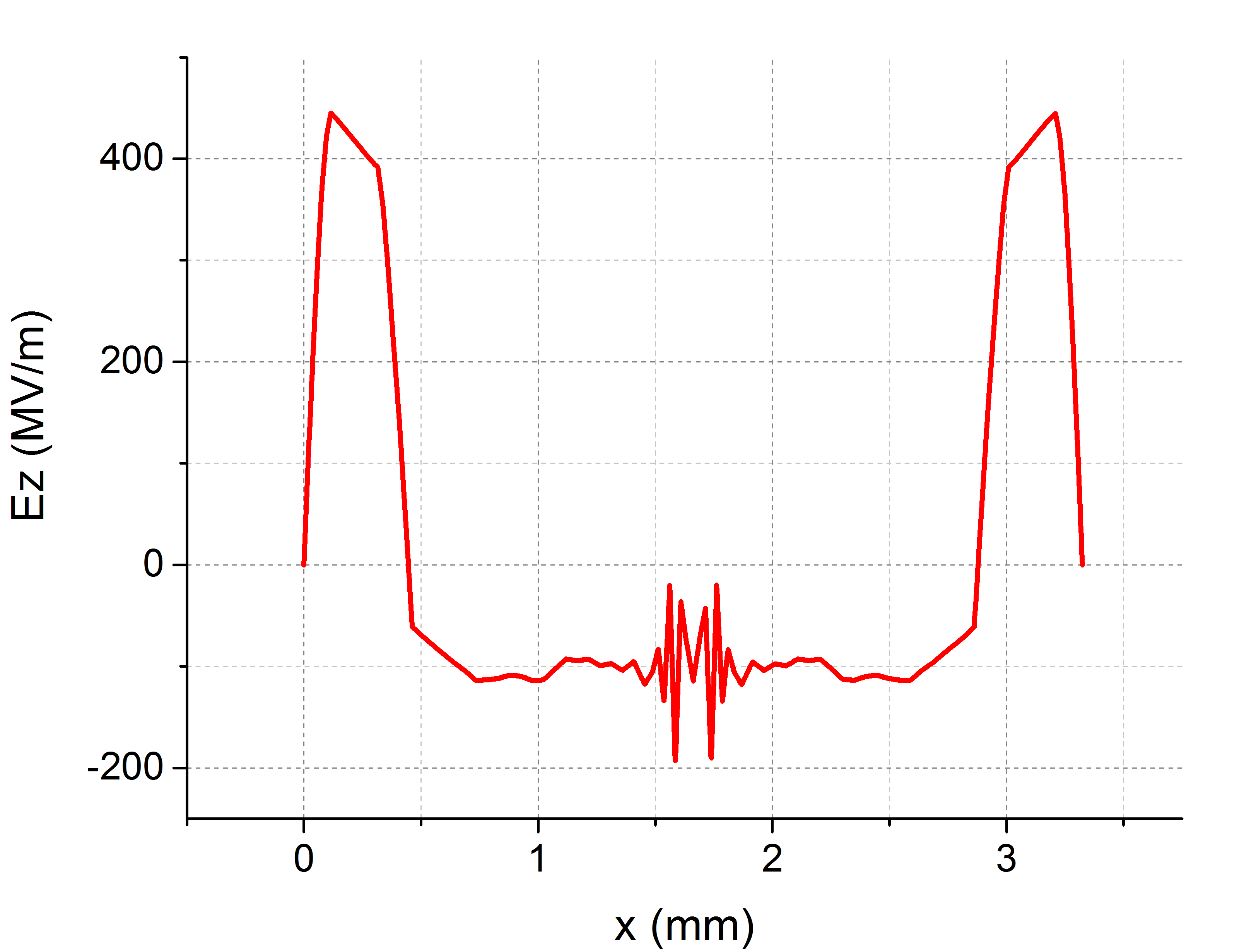}
  \caption{Plot of $E_z(x, y=0, z=0.5L)$,  $t=1.22\,ns$  vs x, from the CST Studio computation (parameters of the simulations are the same in \Figref{Fig:05} except for number of injected bunches equal to 144), revealing a wakefield profile like that of the $LSM_{41}$ mode shown in \Figref{Fig:03}.  }\label{Fig:06}
\end{figure}

A profile of the $E_z$ field inside the resonator after injection of the last drive bunch is displayed in \Figref{Fig:06}. The profile is similar to that of the $LSM_{41}$ mode shown in \Figref{Fig:03}, indicating that a particle placed there will move under the influence of that mode and the system will enjoy an elevated transformer ratio.

This CST Studio simulation takes the wakefield to be perfectly reflected from the ends of the resonator.  However, this can allow the accumulation of radiation from unwanted wakefield modes that have different frequencies as well as broadband radiation emitted from the injection of the drive bunches into the waveguide~\cite{ONI2002PRE}.  The use of tuned Bragg dielectric multilayer mirrors would allow only the wanted $LSM_{41}$ mode to be reflected from the ends of the resonator.

In the present study using CST Studio we can examine the shape of a bunch just  before it exits the resonator (parameters as specified in Table I) at the time when the resonator wakefield is largest.  We have found no deterioration in the shape of the fortieth bunch.  In practice, an actual DWR would be much longer and should have a large but finite Q~\cite{Oni2006AAC-12}, so this result should be revisited.

\begin{figure}
  \centering
  \includegraphics[width=0.48\textwidth]{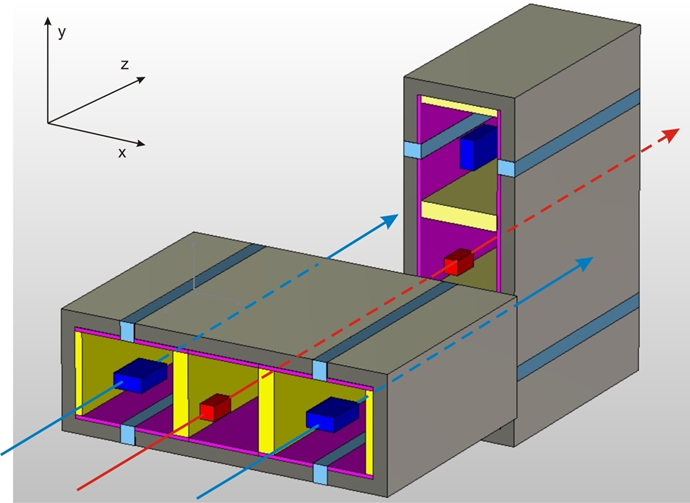}
  \caption{Showing alternating DWR units oriented by a 90 degree rotation in order to dynamically stabilize the drive bunch (in red, traversing the two units)~\cite{Marshall2008AAC-13}.  Four accelerated witness bunches are shown in blue that can be accelerated in such a sequence of resonators.  The light blue slots allow for additional pumping of the resonator and mode control.  The resonator end walls are removed here.  (For clarity, all three channels in this drawing are shown to be the same width, which is not to be the case in practice.)}\label{Fig:07}
\end{figure}

The choice of the best drive bunch shape is open to further study.  A charge profile that is flattened along the y-axis (see \Figref{Fig:01}) and extends a half-$LSM_{41}$-period along the z-axis might be suitable.  However, a small circular cross-section “point-like” drive bunch may also be used, in which case, if the orientation of the tall resonators along the beamline is alternated by $+/-90^o$ around the z-axis, it should be possible to obtain dynamical stabilization of the drive bunch train, because it is found that when expanding the forces about a small region including the z-axis in the drive channel at the drive bunch location, there results transverse forces $F_x = -F_y$, a quadrupole-like effect~\cite{Marshall2008AAC-13} that obtains despite the different height and width dimensions of the resonator.  \Figref{Fig:07} shows how this might be implemented to obtain dynamical stability for the drive bunch.

\section{Conclusions}

A two-beam three channel DWA has been described and shown to be a promising accelerator component for electrons and positrons, combining high gradient operation and elevated transformer ratio with the possibility of improved bunch stability when configured as a resonator.  Additional computational and experimental work, together with investigation of how well dielectrics can survive in an accelerator environment~\cite{Shchelkunov2016NIMA}, are needed to establish the utility of this concept.

\bibliography{Bibliography}	
\end{document}